\documentclass[useAMS,usenatbib]{mn2e}

\usepackage{graphicx,amssymb,url}

\title[Significant radio and $\gamma$-ray correlated variability in
{\it Fermi} bright blazars]{Detection of significant cm to sub-mm band
  radio and $\gamma$-ray correlated variability in {\it Fermi} bright
  blazars} \author[L. Fuhrmann et
al.]{L. Fuhrmann$^{1}$\thanks{E-mail: lfuhrmann@mpifr-bonn.mpg.de},
  S. Larsson$^{2}$, J. Chiang$^{3}$, E. Angelakis$^{1}$,
  J. A. Zensus$^{1}$, I. Nestoras$^{1}$, \newauthor
  T. P. Krichbaum$^{1}$, H. Ungerechts$^{4}$, A. Sievers$^{4}$,
  V. Pavlidou$^{1}$, A. C. S. Readhead$^{5}$,
  \newauthor W. Max-Moerbeck$^{5}$, T. J. Pearson$^{5}$\\
  $^{1}$Max-Planck-Institut f\"ur Radioastronomie, Auf dem H\"ugel 69, D-53121 Bonn, Germany\\
  $^{2}$Oskar Klein Centre, Department of Astronomy, Stockholm University, AlbaNova, SE-10691 Stockholm, Sweden\\
  $^{3}$W. W. Hansen Experimental Physics Laboratory, Kavli Institute
  for Particle Astrophysics and Cosmology, \\
  Department of Physics and SLAC National Accelerator Laboratory, Stanford University, Stanford, CA 94305, USA\\
  $^{4}$Institut de Radio Astronomie Millim\'etrique, Avenida Divina Pastora 7, Local 20, 18012 Granada, Spain\\
  $^{5}$California Institute of Technology, Pasadena, CA 91125, USA}
\begin{document}

\date{November 20, 2013}

\pagerange{\pageref{firstpage}--\pageref{lastpage}} \pubyear{2013}

\maketitle

\label{firstpage}

\begin{abstract}
  The exact location of the $\gamma$-ray emitting region in blazars is
  still controversial. In order to attack this problem we present
  first results of a cross-correlation analysis between radio 
    (11\,cm to 0.8\,mm wavelength, F-GAMMA program) and $\gamma$-ray
    (0.1--300\,GeV) $\sim$\,3.5 year light curves of 54 {\it
      Fermi}-bright blazars. We perform a source stacking analysis and
    estimate significances and chance correlations using mixed source
    correlations.  Our results reveal: (i) the first highly
    significant multi-band radio and $\gamma$-ray correlations (radio
    lagging $\gamma$ rays) when averaging over the whole sample, (ii)
    average time delays (source frame: 76\,$\pm$\,23 to
    7\,$\pm$\,9\,days), systematically decreasing from cm to mm/sub-mm
    bands with a frequency dependence
    $\tau_{\mathrm{r,\gamma}}(\nu)\propto\nu^{-1}$, in good agreement
    with jet opacity dominated by synchrotron self-absorption, (iii) a
    bulk $\gamma$-ray production region typically located
    within/upstream of the 3\,mm core region
    ($\tau_{\mathrm{3mm,\gamma}}=12\pm8$\,days), (iv) mean distances
    between the region of $\gamma$-ray peak emission and the radio
    ``$\tau=1$ photosphere'' decreasing from $9.8\pm3.0$\,pc (11\,cm)
    to $0.9\pm1.1$\,pc (2\,mm) and $1.4\pm0.8$\,pc (0.8\,mm), (v)
    3\,mm/$\gamma$-ray correlations in 9 individual sources at a
    significance level where one is expected by chance (probability:
    $4\times 10^{-6}$), (vi) opacity and ``time lag core shift''
    estimates for quasar 3C\,454.3 providing a lower limit for the
    distance of the bulk $\gamma$-ray production region from the
    supermassive black hole (SMBH) of $\sim$\,0.8--1.6\,pc, i.e.  at
    the outer edge of the Broad Line Region (BLR) or beyond. A 3\,mm
    $\tau=1$ surface at $\sim$\,2--3\,pc from the jet-base (i.e. well
    outside the ``canonical BLR'') finally suggests that BLR material
    extends to several pc distances from the SMBH.
 
 \end{abstract}


\section{Introduction} Since the era of the Energetic Gamma-ray
Experiment Telescope (EGRET) on-board the Compton Gamma-ray
Observatory, the relation between the $\gamma$-ray and radio emission
in Active Galactic Nuclei (AGN) has been intensively discussed. In
particular, the location of the $\gamma$-ray production and
dissipation region in AGN jets is still a matter of active debate --
recently re-activated and intensified thanks to the Large Area
Telescope (LAT) on board the {\it Fermi Gamma-ray Space Telescope}
({\it Fermi}). The LAT is a pair-conversion $\gamma$-ray telescope
sensitive to photon energies from about 20\,MeV up to
$>$\,300\,GeV. Due to its unprecedented sensitivity and all-sky
monitoring capabilities, {\it Fermi}/LAT is providing for the first
time $\gamma$-ray light curves and spectra resolved at a variety of
time scales for a large number ($\sim\,10^{3}$) of AGN since its
launch in 2008
\citep[e.g.][]{2010ApJ...722..520A,2011ApJ...743..171A}.

Different theoretical models and observational findings suggest
different locations of the $\gamma$-ray emitting region, either at (i)
small distances from the central supermassive black hole (SMBH),
i.e. inside the Broad Line Region (BLR, sub-parsec) or even within a
few 100 Schwarzschild radii, very close to the accretion disk
\citep[e.g.][]{1995ApJ...441...79B} or (ii) at larger distances,
e.g. in regions of radio shocks, shock-shock interaction, in various
jet layers or turbulent cells parsecs downstream of the jet
\citep[e.g.][]{1995A&A...297L..13V,2010arXiv1005.5551M,2012A&A...537A..70S,2013arXiv1304.2064M}.
The knowledge of the $\gamma$-ray emission location, however, is of
great importance for any model trying to explain the origin of the
processes responsible for bulk $\gamma$-ray photon production and
energy dissipation
\citep[e.g.][]{2009MNRAS.397..985G,2014ApJ...782...82D}. In leptonic
models, for instance, the exact location of the dissipation region
constrains the origin of the main seed photon fields available for
Inverse Compton (IC) up-scattering to high energies, i.e. either
accretion disk/BLR/jet synchrotron photons ($\lesssim$\,1\,pc) or dust
torus and/or jet synchrotron photons ($\gtrsim$\,1\,pc).

Observationally, several findings disfavor the ``large distance''
scenario, for instance: (i) rapid ($\le$\,hours) MeV/GeV variability
observed in a few sources
\citep[e.g.][]{2010MNRAS.405L..94T,2010MNRAS.408..448F,2013A&A...557A..71R}
suggests ultra-compact emission regions and, assuming that the
emission region is taking up the entire jet cross-section in a conical
jet geometry, a location not too far from the central engine; (ii) the
high-energy spectral breaks observed by {\it Fermi} have been
interpreted as $\gamma$-ray photo-absorption via He\,II Lyman
recombination in the BLR
\citep[][]{2010ApJ...717L.118P,2011MNRAS.417L..11S}, (iii) SED
modeling can often describe well the high energy emission within
leptonic scenarios by external Compton scattering of seed photons from
the BLR and/or accretion disk \citep[e.g.][]{2010ApJ...714L.303F}.

On the other hand, detailed multi-wavelength studies of single sources
including cross-band (radio, optical, X-ray, $\gamma$-ray, and
polarisation) and relative timing analysis of outbursts and{\bf /or}
VLBI component ejection/kinematics suggest relativistic shocks,
shock-shock interaction and/or multiple jet regions on pc scales as
sites of the $\gamma$-ray emission
\citep[e.g.][]{2001ApJ...556..738J,2010ApJ...710L.126M,2010ApJ...715..362J,
  2011ApJ...735L..10A,2012A&A...537A..70S,2013A&A...552A..11R,2013MNRAS.428.2418O,2013MNRAS.436.1530R}.
For instance, the joint occurrence of a $\gamma$-ray flare and an
optical polarization position angle swing observed in 3C\,279 provides
evidence for co-spatial emission regions along a curved trajectory at
a significant distance from the central engine
\citep[][]{2010Natur.463..919A}.  Similarly, joint $\gamma$-ray and
mm-band flares and (mm/optical) polarization peaks along with jet
kinematics also suggest co-spatial emission regions many parsecs
downstream of the jet in OJ\,287 \citep[][]{2011ApJ...726L..13A}.
Finally, rapid variability on time scales of minutes in the few
hundred GeV to TeV energy range can not be produced within the BLR due
to high pair production $\gamma$-ray opacity
\citep*[e.g.][]{2009ApJ...703.1168B,Tavecchio:2012vn}.

\citet{2012ApJ...758L..15D} presented a new method to locate the
energy dissipation region via the energy dependent decay times of
flares in the different cooling regimes of the BLR (Klein-Nishina
regime) and the pc-scale molecular torus region (Thomson
regime). However, this method is limited to the most powerful
$\gamma$-ray events providing enough photon statistics to detect
significant differences in {\it Fermi} light curves at two different
energy bands.  Alternatively, detailed multi-wavelength and
cross-correlation studies of large samples are capable of providing
additional constraints on the location of the $\gamma$-ray emitting
region. For instance, different studies aim at detecting time delays
between $\gamma$ rays and 15\,GHz radio single-dish as well as 
long-term VLBI data of large samples
\citep[e.g.][]{2010ApJ...722L...7P,2013arXiv1303.2131M}, indicating
that cm-band radio flares are generally delayed w.r.t. $\gamma$ rays
\citep[see also][]{2009ApJ...696L..17K}. Parsec-scale distances have
been inferred from delays of $\gamma$-ray peak emission w.r.t. 37\,GHz
radio flare onsets in a sample of sources monitored by the Mets\"ahovi
group \citep[][]{2011A&A...532A.146L}, in-line with earlier results of
similar studies conducted during the EGRET era
\citep[e.g.][]{2003ApJ...590...95L}.

Here, we present the first results of a cross-correlation analysis of
a larger blazar sample based on multi-frequency radio (cm, mm and
sub-mm wavelengths) light curves obtained by the F-GAMMA program
\citep[e.g.][]{2007AIPC..921..249F,fuhrmann2013} and $\sim$\,3.5 year
$\gamma$-ray light curves of 54 {\it Fermi}-bright blazars. The study
aims at (i) establishing statistically significant correlations
between the radio and $\gamma$-ray bands in a sample average sense by
estimating correlation significances and chance correlations using
mixed source correlations as well as a cross-correlation stacking
analysis, and (ii) further constraining the location of the $\gamma$-ray
emitting region in these sources.  The paper is structured as
follows: In Sect.~\ref{data} the sample and data sets are
introduced. Sect.~\ref{DCCF_analysis} describes the applied
cross-correlation methods and analysis, whereas Sect.~\ref{results}
and \ref{discussion} present and discuss the results.  A summary and
concluding remarks are given in Sect.~\ref{conclusions}.

\begin{figure*}
\centering
\includegraphics[width=176mm,angle=0]{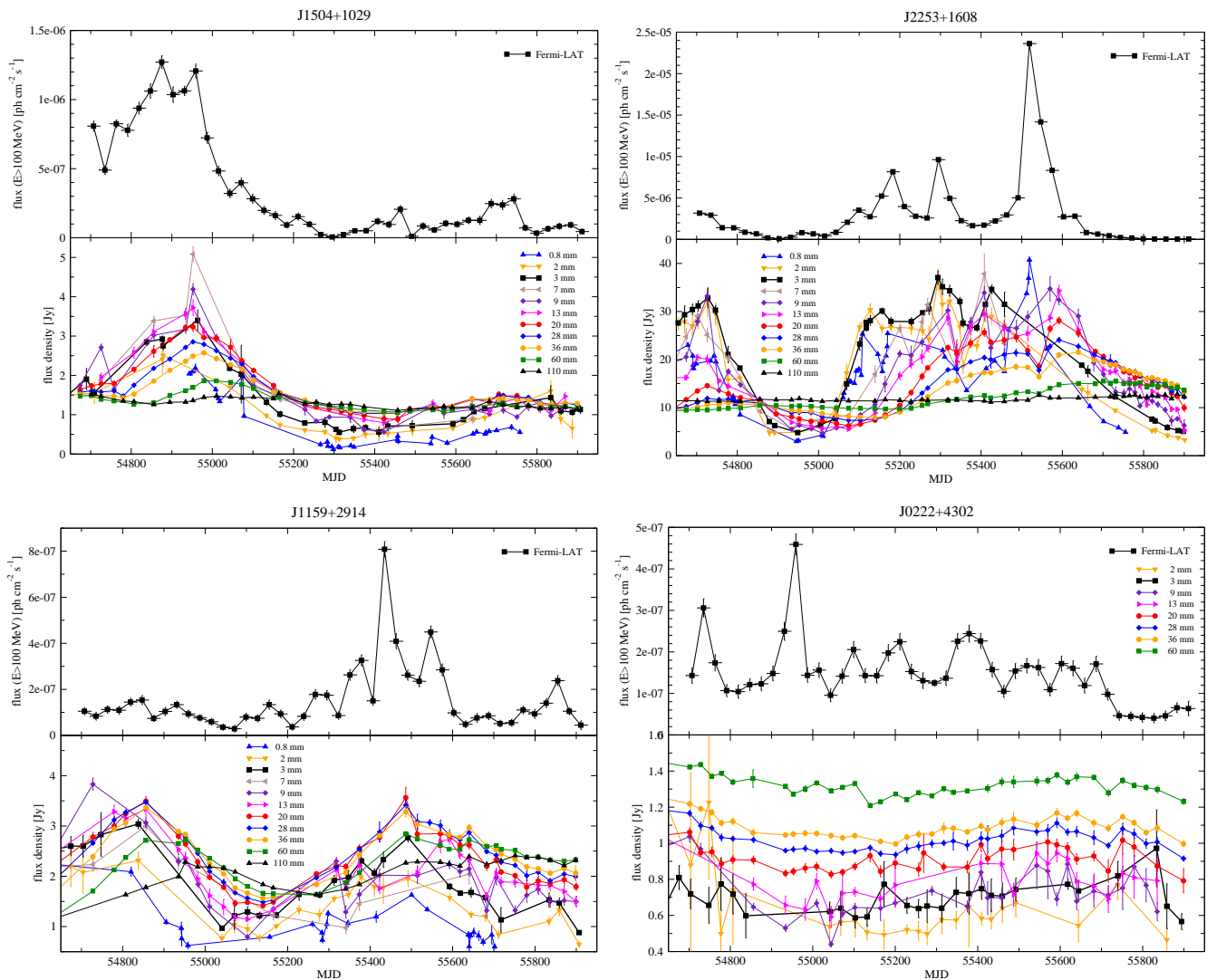}
\caption{The $\gamma$-ray (top) and radio (bottom) light curves (flux
  vs. modified julian date, MJD) for four selected, bright
  $\gamma$-ray sources of the studied sample: J1504+1029
  (PKS\,1502+106, top left; redshift: 1.84), J2253+1608
  (3C\,454.3, top right; redshift: 0.86), J1159+2914 (4C\,29.45,
  bottom left; redshift: 0.73) and J0222+4302 (3C\,66A, bottom
  right; redshift: 0.44). The top two and bottom left sources
  demonstrate cases of possible correlations between both bands,
  whereas no correlated variability is evident for J0222+4302 (bottom
  right).}
\label{fgamma_lat_LCs}
\end{figure*}

\section{The source sample and data sets}\label{data}

\begin{table}
\centering
\caption{Summary of the different wave/energy bands used in
the current analysis.}
\begin{tabular}{llll}
\hline
Facility            & Band             & Frequency & Energy \\
(F-GAMMA)           & [mm]             & [GHz]     & [GeV] \\
\hline
Effelsberg 100-m    & 110,\,60,\,36     & 2.64,\,4.85,\,8.35   & --\\
                    & 28,\,20,\,13      & 10.45,\,14.6,\,23.05 &--\\
                    & 9,\,7             & 32.0,\,43.0         &--\\
IRAM 30-m           & 3,\,2             & 86.2,\,142.3 &--\\
APEX 12-m           & 0.8              & 345 &--\\
\hline
{\it Fermi}/LAT     & --               & --  & 0.1--300 \\
\hline
\end{tabular}
\label{radio_bands}
\end{table}

\subsection{The F-GAMMA program: cm to sub-mm band light curves}\label{fgamma_data}
The cm, mm and sub-mm band radio data used for the current study have
been collected in the framework of the {\sl Fermi} related F-GAMMA
monitoring program
\citep[][]{2007AIPC..921..249F,2010arXiv1006.5610A,fuhrmann2013}.
Since 2007, the F-GAMMA program has been monitoring contemporaneously
the total flux density, polarisation and spectral evolution of about
60 {\it Fermi} blazars at three radio observatories, enabling detailed
AGN studies of broad band variability, emission processes as well as
the radio/$\gamma$-ray connection. The overall wavelength range spans
110 to 0.8\,mm (2.64 to 345\,GHz) using the Effelsberg (EB) 100-m,
IRAM 30-m (at Pico Veleta, PV) and APEX (Atacama Pathfinder
EXperiment) 12-m telescopes at a total of 11 bands (see Table
\ref{radio_bands}). The monthly observations at EB and PV are
performed quasi-simultaneously (typically within days) and in a highly
synchronised manner together with the more general flux monitoring
conducted at the IRAM 30-m telescope. APEX sub-mm observations are
performed for 25 F-GAMMA sources in addition to a sample of
interesting southern hemisphere {\it Fermi}-detected AGN not
observable from the EB and PV sites.

The Effelsberg measurements were conducted with cross-scans using the
secondary focus heterodyne receivers at 8 wavebands between 110 and
7\,mm wavelength (2.64 to 43.00\,GHz, see Table \ref{radio_bands}).
The IRAM 30-m observations were carried out with calibrated
cross-scans using the ``B'' and ``C'' SIS (until March 2009) and EMIR
(Eight Mixer Receiver) heterodyne receivers operating at 3 and 2\,mm
wavelength (86.2 and 142.3\,GHz). Finally, the Large Apex Bolometer
Camera (LABOCA) array was used at APEX operating at a wavelength of
0.87\,mm (345\,GHz). In the data reduction process for each station,
pointing offset, gain-elevation, atmospheric opacity and sensitivity
corrections have been applied to the data. The details of the program,
observations and data reduction are described in \citet{fuhrmann2013},
\citet{nestoras2013} and \citet{2012arXiv1206.3799L} \citep[see
also][Angelakis et al. in
prep.]{2008A&A...490.1019F,2010arXiv1006.5610A}. Example light curves
of four selected sources (J0222+4302, J1159+2914, J1504+1029 and
J2253+1608) including all radio bands are shown in
Fig.~\ref{fgamma_lat_LCs}. The data of J0222+4302 (3C\,66A) at 110\,mm
wavelength are affected by the close-by radio galaxy 3C\,66B and thus
have been omitted.

\subsection{The source sample}\label{sample}
The present study is focusing on the sources observed at radio bands
by the F-GAMMA monitoring program. With a total of about 90
AGN/blazars ever observed since January 2007, these sources constitute
a sample of well known, frequently active and bright blazars ($\delta
> -30^\circ$) for detailed studies of the {\it most prominent}
behavior of the {\it brightest} $\gamma$-ray--loud blazars.

For the particular analysis presented here we selected a sub-sample
from the above F-GAMMA sources according to the following criteria:
(i) {\it Fermi}-detection: presence in the 1FGL catalog, (ii) ``best
suitable'' radio light curves: sources with the best frequency and
time coverage that are well sampled over the considered {\it
  Fermi} time period of about 3.5 years, (iii) presence of radio
variability: sources showing significant variability on the basis of a
$\chi^2$-test.

This selection results in a sub-sample of 54 sources comprised of 35
Flat-Spectrum Radio Quasars (FSRQs), 18 BL Lacertae objects (BL\,Lacs)
and one Narrow-Line Seyfert 1 galaxy. Table \ref{source_sample}
presents the complete list of selected sources. At the sub-mm band the
selected sub-sample is slightly different due to the different source
sample observed at the APEX telescope. In this case, a total of 38
sources have been selected, including (i) 23 sources of the above
sub-sample of 54 sources and (ii) 15 additional, southern hemisphere
{\it Fermi}-detected AGN also satisfying the above criteria. The
latter are given at the bottom of Table \ref{source_sample}.

Given our source selection and the resulting statistical
incompleteness of the studied sample, we note that the results
presented in the following may not be representative of the
AGN/blazar population in its entirety.

\begin{table}
\centering
\caption{List of selected sources included in the current analysis (see text for details).}
\begin{tabular}{llll}
\hline
Source  & 2FGL name    & other name  & type \\
\hline
J0050$-$0929 &  J0050.6$-$0929  &  PKS\,0048$-$097  & BL\,Lac\\
J0102+5824   &  J0102.7+5827    &  TXS\,0059+581    & FSRQ\\
J0136+4751   &  J0136.9+4751    &  OC\,457          & FSRQ\\
J0217+0144   &  J0217.9+0143    &  PKS\,0215+015    & FSRQ\\
J0222+4302   &  J0222.6+4302    &  3C\,66A          & BL\,Lac\\
J0237+2848   &  J0237.8+2846    &  4C\,+28.07       & FSRQ\\
J0238+1636   &  J0238.7+1637    &  AO\,0235+164     & BL\,Lac\\
...          &                  &                   &  \\
\hline
\end{tabular}
{\footnotesize Note: this table is available in its entirety as online material.}
\label{source_sample}
\end{table}

\subsection{{\it Fermi} $\gamma$-ray light curves} 

The {\it Fermi} $\gamma$-ray light curves for the studied sample have
been produced in a pipeline fashion using time boundaries to best
match the radio light curves: a 28-day binning starting on August 15,
2008 and ending on January 26, 2012 was used. The choice of 28-day
binning was primarily driven by the predetermined cadence of the
F-GAMMA radio light curves (about one month). Furthermore, this choice
is also a trade-off between time resolution and good signal-to-noise
ratio in each time bin for low $\gamma$-ray flux states and/or weaker
sources in the sample.
  
The source model for each {\it region-of-interest} (ROI) containing
the target sources included nearby point sources, determined from the
Second {\it Fermi}/LAT catalog \citep[2FGL,][]{2012ApJS..199...31N},
and the standard Galactic and isotropic diffuse emission models (two
year P7V6 models\footnote{see also:
  \url{http://fermi.gsfc.nasa.gov/ssc/data/access/lat/BackgroundModels.html}}).
The latter component includes contributions from unresolved
extragalactic emission and any residual charged particle
backgrounds. A maximum likelihood analysis of each ROI was performed
with ScienceTools version 09-26-00, and the P7SOURCE\_V6 instrument
response functions (IRFs). The fluxes and photon spectral indices were
fit for each target source assuming a single power-law over the energy
range 0.1--300\,GeV. Examples of the $\gamma$-ray light curves are
shown in Fig.~\ref{fgamma_lat_LCs}. We note that due to the 28-day
binning interval, more rapid $\gamma$-ray flares and variability on
time scales of hours/days to a few weeks is smoothed out and not
resolved with our data sets.

In total, {\it Fermi} $\gamma$-ray light curves were produced for 131
sources. In addition to our studied sample of 54 sources, we included
77 reference blazars with good quality {\it Fermi} light curves that
were used for estimating correlation significances, as described in
Sect. \ref{DCCF_analysis}. The reference sources were chosen among the
brightest blazars in the Second {\it Fermi}/LAT AGN catalog
\citep[2LAC,][]{2011ApJ...743..171A}, including sources of each blazar
type (FSRQs; low, intermediate and high synchrotron peaked
BL\,Lacs). Their $\gamma$-ray light curves were produced in the same
way as the light curves for our sample of 54 target sources.

\section{Cross-correlation analysis}\label{DCCF_analysis}

\subsection{The DCCF method}\label{dccf_method}

\begin{figure}
\centering
\includegraphics[width=85mm, angle=0]{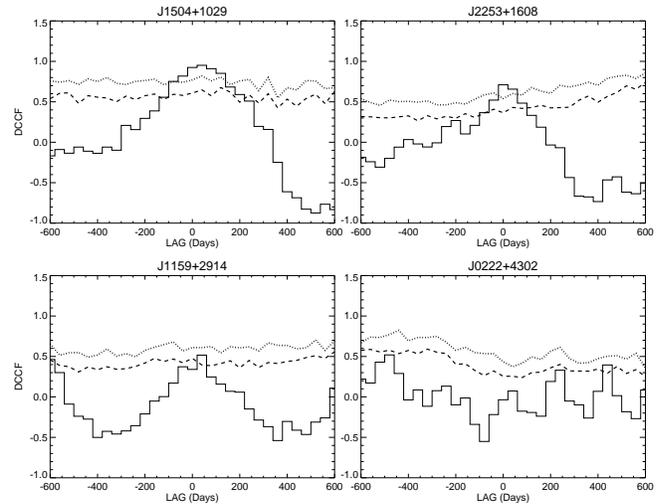}
\caption{3\,mm/$\gamma$-ray DCCFs (observers frame) of the single
  sources also shown in Fig. \ref{fgamma_lat_LCs}, namely J1504+1029
  (top, left), J2258+1608 (top, right), J1159+2914 (bottom, left) and
  J0222+4302 (bottom, right), with 99\,\% (dotted lines) and 90\,\%
  (dashed lines) significance levels superimposed. Only the top two
  cases show significant correlations above 99\,\% significance.}
\label{DCCF_examples}
\end{figure}

In order to search for possible correlations between the $\gamma$-ray
and radio light curves, we use a cross-correlation analysis. For two
discrete, evenly sampled light curves, $x(t_i)$ and $y(t_i)$, the
Cross-Correlation Function (CCF) as function of time lag $\tau$ is
defined as

\begin{equation}\label{eq:ccf}
CCF(\tau)=\frac{1}{N}\sum_{i=1}^{N}{\frac{[x(t_{i})-\bar{x}][y(t_{i}-\tau)-\bar{y}]}{\sigma_{x}\sigma_{y}}}\,\,,
\end{equation} 
with $\bar{x}$, $\sigma_{x}$ and $\bar{y}$, $\sigma_{y}$ the mean and
standard deviation of $x(t_{i})$ and $y(t_{i})$, respectively. Given
the uneven sampling of the light curves considered here, we in
particular use a Discrete Cross-Correlation Function (DCCF) analysis
\citep{1988ApJ...333..646E}, where in contrast to linear interpolation
methods, the contribution to the CCF is calculated only using the
actual data points. Each pair of points, one from each of the two
light curves, then provides one correlation value at a lag
corresponding to their time separation. For two light curves with N
and M data points respectively, this gives an Unbinned Cross
Correlation Function (UCCF)

\begin{equation}\label{eq:uccf}
UCCF_{ij} = \frac{(x_i-\bar{x})(y_j-\bar{y}) }{\sigma_x \sigma_y}\,\,.
\end{equation}
The DCCF is then obtained by averaging the UCCF in time lag bins. For
light curves exhibiting stationary variability, the mean and variance
is the same for different parts of the data, except for statistical
fluctuations.  For non-stationary light curves the mean and variance
may exhibit larger fluctuations and in order to take this into account
it is advantageous to calculate a new mean and variance for each time
lag bin, using only the data points that contribute to the DCCF at
that lag \citep[see][]{1994PASP..106..879W}.  With this definition the
value of the DCCF is identical to Pearson's r-statistic.  Values can
vary between $-1$ and $+1$. A positive value implies correlated
variability and a negative value corresponds to an anti-correlation.
A DCCF peak at some time lag $\tau$ implies, in case it is significant
(see Sect.~\ref{significance}), correlated variability with an average
time shift $\tau$ between the two time series. In all our analysis
positive lag denotes $\gamma$-ray leading radio. For each source,
DCCFs were calculated between the $\gamma$-ray light curve and each of
the radio light curves (0.8, 2, 3, 9, 7, 13, 20, 28, 36, 60 and
110\,mm wavelength).  Examples of 3\,mm/$\gamma$-ray DCCFs for four
single sources are shown in Fig.~\ref{DCCF_examples}. For instance,
the DCCF of J1504+1029 (top left) shows a clear positive peak in the
DCCF with amplitude close to one at a small positive time lag $\tau$
indicating correlated variability with 3\,mm radio lagging behind
$\gamma$ rays by a certain time lag. In contrast, J0222+4302 (bottom
left) shows no prominent, positive or negative DCCF peak. Hence, no
correlation between the two bands is detected over the time period of
about 3.5 years.

To estimate the location and time lag of each DCCF peak, we fit a
Gaussian function over a lag range of 200 days for the short
wavelength bands ($\le$\,3\,mm) to 300 days for the longer wavelengths
($\ge$\,7\,mm). Uncertainties were estimated by a model independent
Monte Carlo method \citep{1998PASP..110..660P} accounting for the
effects of measurement noise and data sampling.  The Monte Carlo run
consisted of a bootstrap selection of a subsample of data points in
each light curve plus injection of white noise with a standard
deviation equal to the error value at each data point. The
uncertainties in correlation time lags are then estimated as the
standard deviations of peak fits to these simulations.

\subsection{Correlation significance: mixed source correlations}\label{significance}

The strength and significance of peaks seen in the DCCF depends on the
stochastic nature of the variability, the correlation properties, the
data sampling, the measurement noise and the total length of the time
series. The latter limits the number of observed events for a given
source which furthermore depends on the source duty cycle and the time
scale of the observed variability. For the data used in the present
study, the correlation significance is primarily limited by chance
correlations, i.e. DCCF peaks coming from physically unrelated
variability in the two different spectral bands. If, for instance, the
two light curves contain flares at times that are unrelated to each
other, a correlation peak will be seen in the DCCF at a lag
corresponding to their time separation even if they are causally
unrelated. If the number of flares or variability features in the
light curves are small these effects can be strong.

In order to estimate the probability that observed correlations are
produced by chance correlations or observational effects we compute
``mixed source correlations''. This is done by correlating the radio
light curve for a given source with all other 130 available
$\gamma$-ray light curves in the same manner as described in Sect.
\ref{dccf_method} and then compare, for each lag bin, the distribution
of these correlation values with the actual DCCF value of the source
at the same lag. Under the assumption that the variability properties
are similar for the different sources, the probability distribution of
the resulting DCCFs reflect the occurrence of spurious correlations.
The assumption of similar variability properties is supported by the
fact that the 131 $\gamma$-ray light curves all show significant
variability and are largely dominated by red-noise like processes as
demonstrated by a power spectral density (PSD) analysis.  The latter
is in good agreement with previous findings of
\cite{2011ApJ...743..171A} using 1 month binned, 2 year {\it Fermi}
light curves of 156 FSRQs and 59 BL\,Lacs (including a large fraction
of our sources).

Further details on the correlation techniques and significance 
estimates are given in \citet{Larsson:2012vu}.

\subsection{DCCF stacking analysis}\label{dccf_stacking_method}

With the present data our analysis is able to reveal significant
correlations for a handful of sources as described in the next
section. However, the sensitivity for the detection of correlations
and their multi-frequency properties as well as the DCCF peak
significance can be largely improved by using the whole source
sample. For that reason we perform a joint, or stacking, analysis and
we do so in two different ways: (i) we simply average the DCCFs of all
the sources, (ii) we first normalize the light curves by dividing with
the mean flux density for each source and then include the
contribution of all correlation data pairs in the computation of an
average DCCF. The two methods give similar, but not identical
results. In particular, the second method gives more weight to sources
with large variability and to source light curves with a higher
density of data points. The correlation significance is again
estimated by mixed source correlations that are stacked in a similar
way as the real source DCCFs. Each mixed source DCCF is only used
once, which results in 130 stacked comparison DCCFs.

In the last part of the analysis we quantify how the correlation
depends on radio wavelength. Here, we also take into account the
cosmological time stretch by scaling time values with a factor of
$1/(1+z)$, such that the computed time lags refer to source rest
frame. We note that in the current work we do not take into account
that time values (time scales and time lags) are additionally modified
(shortened) by relativistic boosting effects (``jet rest
frame''). This aspect will be addressed in a subsequent analysis
(Larsson et al. in prep.).
 
The same 54 sources (mean redshift: 0.9) were used for
each one of the 2--110\,mm/$\gamma$-ray band combination, except for
one source (J1626-2948) which was excluded from the 7--110\,mm band
DCCFs due to poor sampling. For the 0.8\,mm DCCF, 38 APEX light curves
were used, out of which 23 sources were also in the F-GAMMA sample
used for the longer wavelengths.

\section{Results}\label{results}

Since the different radio bands are usually correlated (both in
sampling and variability) and radio/$\gamma$-ray correlations are
expected to be more pronounced towards shorter wavelengths
\citep[e.g.][]{fuhrmann2013}, we only consider one radio band (3\,mm)
in order to establish the correlation significance. We choose the
3\,mm band as being our best data set at short wavelengths in terms of
sampling and measurement noise. We first present the results of the
stacking analysis in Sect. \ref{sign_stacking}, whereas the most
significant single-source correlations of our analysis are reported in
Sect. \ref{results_single_sources}. In
Sect. \ref{stacked_DCCF_results} we investigate how the average
correlation depends on radio wavelength using the stacked DCCFs.

\subsection{Correlation significance in stacking analysis}\label{sign_stacking}

\begin{figure}
\centering
\includegraphics[width=82mm, angle=0]{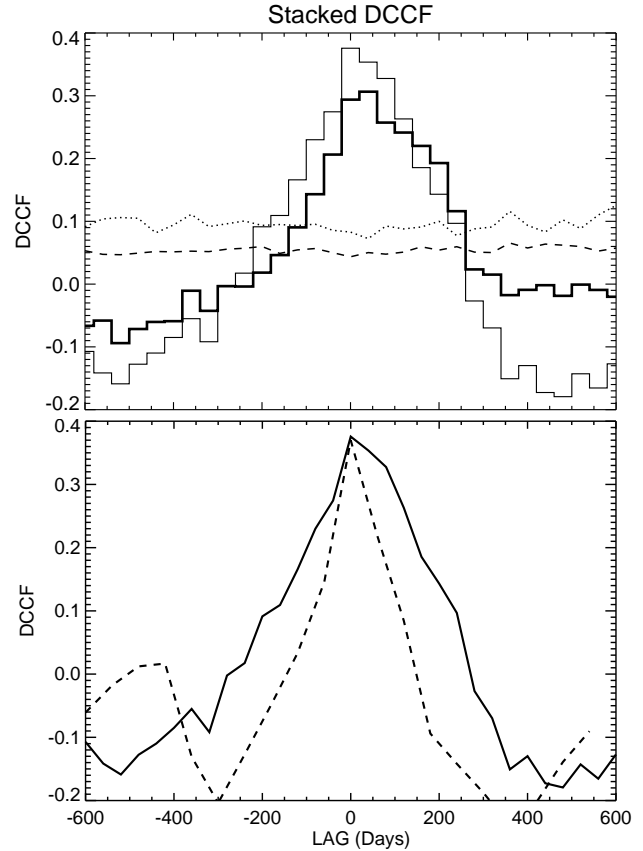}
\caption{Averaged 3\,mm/$\gamma$-ray DCCFs obtained from the stacking
  analysis. Top: DCCFs (observers frame) of both averaging methods are
  shown (bold line: direct averaging of DCCFs, see text). 99 (dotted
  line) and 90\,\% (dashed line) significance levels for the direct
  averaging method are superimposed demonstrating the detection of
  highly significant correlations. Bottom: comparison of the stacked
  DCCFs obtained with (source frame, dashed line) and without
  (observers frame, solid line) redshift correction.}
\label{stacked_DCCF}
\end{figure}

A 3\,mm/$\gamma$-ray source averaged DCCF was calculated with each one
of the two stacking methods described in Section
\ref{dccf_stacking_method}.  Both methods give a highly significant
detection of correlated variability as can be seen in the top panel of
Fig. \ref{stacked_DCCF}, where the two DCCFs are shown together with
90 and 99\,\% significance levels as estimated from the mixed source
correlations.  The second stacking method (building the DCCF by adding
data point pairs) results in a correlation peak DCCF$_{{\mathrm max}}$
of 0.38 compared to 0.31 for the direct averaging of the individual
source DCCFs. In both cases the DCCF peak (average for lag $-$100 to
$+$100 days) is more than 8 times higher than the strongest of the
mixed source DCCF used for comparison. We note that even after
removing the 12 sources with the strongest correlations from the
analysis, the DCCF correlation is still highly significant. In this
case the average for the lag range $-$100 to $+$100\,days still
exceeds the strongest corresponding mixed source DCCF by a factor of
4. This demonstrates that the overall correlation is not restricted to
or dominated by just a small fraction of the sources.

The correlation peak in the stacked DCCF of Fig. \ref{stacked_DCCF}
(top) is broad and extends over a positive {\it and} negative lag
range of several hundred days. This width is partly the result of the
distribution of correlation lags among our sources, including the
redshift effect (see the bottom panel of Fig. \ref{stacked_DCCF}) and
the possible presence of multiple lags in individual sources, but
mostly it is an effect of the variability time scale (see also
Sect. \ref{stacked_DCCF_results}). If the length of a $\gamma$-ray
flaring period is longer than the time delay of the radio flare onset,
the later part of the $\gamma$-ray flare will correlate with the
beginning of the radio flare at a negative time lag.  This is the main
reason why the DCCF peak extends to negative lags -- consistent with
our findings of single sources not showing significant negative time
lags (see Sect. \ref{results_single_sources}).

\subsection{Single sources: light curves and 3\,mm/$\gamma$-ray DCCFs}\label{results_single_sources} 

The example light curves presented in Fig.~\ref{fgamma_lat_LCs}
demonstrate the typical behavior seen in the studied sample of 54
sources.  A visual inspection of all light curves shows that (i)
strong flux density outbursts (time scales of months) and extended
periods of activity (months to 1--2 years) occur at both radio and
$\gamma$ rays, although single $\gamma$-ray events usually appear to
be more rapid, (ii) the flaring activity is often characterised by
significant sub-structure with faster sub-flares superimposed, in
particular at $\gamma$-rays, (iii) the flaring activity often seems to
happen quasi-simultaneously at both bands, i.e. during periods of
$\gamma$-ray activity the radio bands are correspondingly in an
increasing or high flux/activity state, (iv) often obvious time lags
(radio lagging) are evident between the peaks of radio and
$\gamma$-ray flares, in particular towards longer radio
wavelengths. The sources J1504+1029 and J2253+1608 shown in
Fig. \ref{fgamma_lat_LCs} are typical examples of the described
characteristics. On the other hand, we identify several sources, like
J0222+4302 (see Fig. \ref{fgamma_lat_LCs}), showing no obvious
correlation. Despite substantial flaring activity at $\gamma$ rays,
there is no obvious corresponding variability at radio
bands. Occasionally we also find (i) strong (factor $\sim$\,3--4)
outbursts in one band with only a very mild ``counterpart'' at the
other spectral band, (ii) very rapid variability and flares in both
bands without obvious, simple ``one-to-one'' correspondences of
events.  Finally, we find evidence that the mm/sub-mm band flux
density rises simultaneously with or even before the $\gamma$ rays in
a few cases.

The examples of 3\,mm/$\gamma$-ray DCCFs for individual sources shown
in Fig.~\ref{DCCF_examples} confirm the impression of correlated
variability in the light curves of the first two cases (J1504+1029 and
J2253+1608): a prominent peak close to zero lag and above our
significance levels is seen. In contrast, no DCCF peak is seen in the
case of J0222+4302. In the case of J1159+2914 we find a prominent DCCF
peak close to zero lag, though well below the 99\,\% significance
level. 

\begin{table}
\centering
\caption{Single sources showing the most significant
  3\,mm/$\gamma$-ray correlations. Source type, estimated time lag
  $\tau_{{\mathrm 3mm,\gamma}}$ (source frame), DCCF peak and 
  $\Delta r_{{\mathrm 3mm,\gamma}}$ are given (see text). Positive lag denotes
  $\gamma$-ray leading radio. For J0050-0929 no
  $\beta_{\mathrm{app}}$ was available to estimate 
  $\theta$ and $\Delta r_{{\mathrm 3mm,\gamma}}$. An estimate for J0238+1636 
  has been omitted (see text).}
 \begin{tabular}{llllc}
 \hline
     Source  & type     & lag       & DCCF$_{max}$  & $\Delta r_{{\mathrm 3mm/\gamma}}$\\
	     &          & [days]    &              & [pc]\\
\hline
J0050$-$0929 &  BL\,Lac & ~$48\pm26$   & $0.88\pm0.09$ & --\\
J0237+2848   &  FSRQ    & ~$40\pm10$   & $0.90\pm0.13$ & ~3.6\\
J0238+1636   &  BL\,Lac & $-4\pm10$    & $1.00\pm0.17$ & -- \\
J0530+1331   &  FSRQ    & ~$10\pm\,\,\,8$   & $0.82\pm0.06$ & ~0.3\\
J1504+1029   &  FSRQ    & ~$14\pm11$   & $0.96\pm0.05$ & ~2.1\\
J1733$-$1304 &  FSRQ    & ~$29\pm26$   & $0.85\pm0.13$ & 15.9\\
J2147+0929   &  FSRQ    & ~$15\pm15$   & $0.82\pm0.09$ & ~0.2\\
J2202+4216   &  BL\,Lac & ~$93\pm16$   & $0.82\pm0.10$ & ~4.7\\
J2253+1608   &  FSRQ    & ~~\,$8\pm12$   & $0.73\pm0.06$ & ~0.9\\
\hline
\end{tabular}
\label{single_source_results}
\end{table}

The 90 and 99\,\% significance levels shown in
Fig. \ref{DCCF_examples} refer to individual time bins in the DCCF and
are estimated from comparisons with the mixed source correlations.
Based on the strong and well defined correlation peak seen in the
3\,mm/$\gamma$-ray stacked DCCF of the whole sample, we compared, for
each individual source, the DCCF level for the lag range from $-$100
to $+$140 days with the corresponding level for the 130 mixed-source
DCCFs. In 5 out of 54 sources the DCCF level exceeds all 130
comparison DCCFs (corresponding to a significance $>$\,99\,\%) and for
another 4 sources the level is exceeded by 1 or 2 comparison DCCFs
(significance $\gtrsim$\,98\,\%). Statistically we would expect about
one case occurring by chance. The chance probability to obtain 9 cases
or more is only $4\times 10^{-6}$. These 9 sources with the most
significant correlations are listed in Table
\ref{single_source_results} along with their estimated
3\,mm/$\gamma$-ray time lag $\tau_{{\mathrm 3mm,\gamma}}$ (source
frame). The estimated lags range between $-4\pm10$ and $93\pm16$\,days
with a mean and median value of 28 and 15 days, respectively.

Radio/$\gamma$-ray light curve cross-correlations and time lags have
been reported previously for a few sources of Table
\ref{single_source_results}. In the case of J2202+4216 (BL\,Lacertae),
\citet[][]{2013MNRAS.436.1530R} estimated a lag of
$\sim$\,120--150\,days (observers frame) with the mm-bands lagging
$\gamma$ rays. As seen in Table \ref{single_source_results}, our
analysis confirms a correlation in this source, though with a shorter
lag of $\sim$\,100\,days (observers frame). We note, however, that the
analysis of \citet[][]{2013MNRAS.436.1530R} covered a significantly
longer time period (up to October 31, 2012) including the latest high
activity period of BL Lacertae during
2012. \citet[][]{2012ApJ...758...72W} \citep[see
also][]{2013ApJ...773..147J} reported a significant correlation with
time lags close to zero using 1.3\,mm SMA and {\it Fermi} $\gamma$-ray
light curves of J2253+1608 (3C\,454.3) up to October 2011
(i.e. comparable to our time range). As seen in
Fig. \ref{DCCF_examples} and Table \ref{single_source_results}, our
analysis confirms this result. The estimated time lags are in good
agreement given the lag uncertainties and the shorter observing
wavelength of the SMA data. \citet[][]{2011ApJ...735L..10A} found
significant correlations between $\gamma$ rays and cm/mm bands for the
prominent flaring activity of J0238+1636 (AO 0235+164) during the
early phase of the Fermi mission (August 2008). The
1.3\,mm/$\gamma$-ray DCCF of these authors is broad with peaks at lags
$\sim$\,0 and $\sim$\,50\,days ($\gamma$-ray leading).  We stress that
due to the 28-day binning interval our $\gamma$-ray light curve does
not resolve the rising part of this flare which adds an additional
systematic uncertainty. Consequently, given the differences in time
range, sampling and time binning with respect to our analysis it is
difficult to make a quantitative comparison of these results. We do
note, however, that there is a qualitative agreement between the lag
estimates of these studies.

As seen in Table \ref{single_source_results}, we do not find sources
exhibiting a significant negative time lag, i.e. no significant case
of 3\,mm leading the $\gamma$ rays is found. As is demonstrated by our
stacking analysis, the low detection rate of significant single-source
correlations does not imply the non-existence of correlations in the
majority of single sources. While several cases like J0222+4302 in
Fig. \ref{DCCF_examples} are present, the low detection rate is
primarily induced by the limited time span of 3.5 years, i.e. the so
far limited statistics and small number of available events for single
sources. Here, longer data trains will significantly improve the
situation. A detailed single-source analysis using 5 year data sets is
in progress (Fuhrmann et al. in prep.). Furthermore, Table
\ref{single_source_results} shows that the current analysis reveals
significant single-source correlations for both FSRQs and BL\,Lacs in
the sample. This motivates future studies of stacked correlations
separated by source type to search for possible differences in
correlation behavior (correlation strength, time lags etc., Larsson et
al. in prep.).

\subsection{Stacked DCCF as a function of wavelength}\label{stacked_DCCF_results} 

\begin{figure}
\centering
\includegraphics[width=83mm,angle=0]{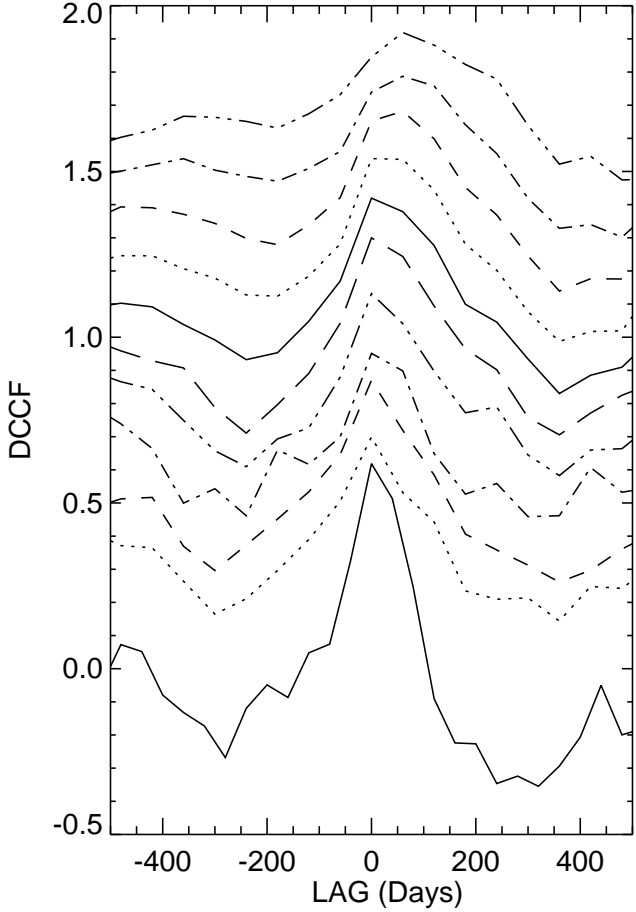}
\caption{Stacked radio/$\gamma$-ray DCCFs (source frame) across all
  radio bands.  From top to bottom are shown: $\gamma$-ray vs. 110,
  60, 36, 28, 20, 13, 9, 7, 3, 2 and 0.8\,mm wavelength. For better
  illustration, the 2\,mm/$\gamma$-ray DCCF has been displaced along
  the y-axis by 0.35 and the longer radio wavelengths ones each by an
  additional shift of 0.15.  Since time values are redshift corrected,
  fewer data points contribute to the stacked DCCF at large lags which
  increases the sensitivity to chance correlations (such as the peaks
  at lags $-$450 and $+$450 days).}
\label{all_stacked}
\end{figure}

Having established the presence of a highly significant
3\,mm/$\gamma$-ray correlation in the stacked analysis we now repeat
the analysis for each of the radio bands.

In Fig. \ref{all_stacked} the average DCCFs for all radio/$\gamma$-ray
combinations are presented. The first thing to be noticed is an
increase in DCCF width towards longer cm-bands. In particular, an
asymmetric DCCF shape is seen with a wing extending to larger radio
lags and becoming more pronounced towards 110\,mm wavelength,
consistent with the successively longer variability time scales and
more extended flare shapes seen at longer cm-bands.  Furthermore, the
correlation peak is close to time lag zero for the shortest
wavelengths and shifts towards larger, positive ($\gamma$-ray leading)
time lags with increasing radio wavelength. Finally, the correlation
peak maxima increase towards the sub-mm band from DCCF$_{{\mathrm
    max}}=0.23\pm0.05$ at 110\,mm to DCCF$_{{\mathrm
    max}}=0.61\pm0.05$ at 0.8\,mm.

\begin{figure}
\includegraphics[width=85mm,angle=0]{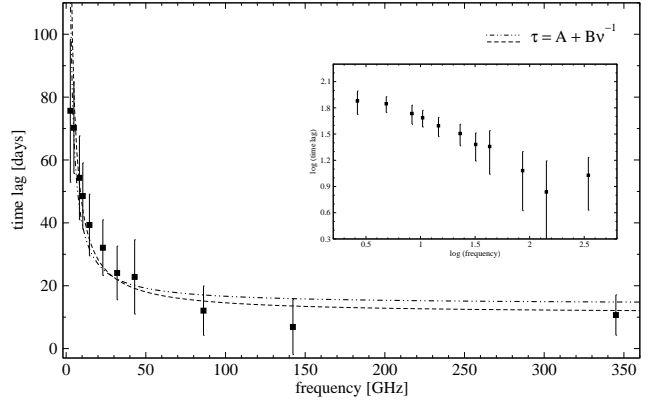}
\caption{Stacked radio/$\gamma$-ray time lags (source frame) vs.
  frequency. Positive lags denote $\gamma$-ray leading. A clear trend
  of decreasing lags towards higher frequencies is evident.  The lines
  represent least-square fits of the form
  $\tau_{\mathrm{r,\gamma}}(\nu)=A+B\,\nu^{-1}$ to all lags
  (dashed-dotted line) as well as omitting the lag of the lowest radio
  frequency (dashed line). For the former case, the fit parameters A
  and B are 14.1 and 257.4, respectively. The inset displays the data
  in a logarithmic representation.}
\label{lag_vs_wavelength}
\end{figure}

The estimated (source frame) time lags with uncertainties are shown in
Fig. \ref{lag_vs_wavelength} as a function of radio frequency. The
average time lag increases smoothly from $7\pm9$\,days at 142\,GHz
(2\,mm) to $76\pm23$\,days at 2.6\,GHz (110\,mm). The errors given in
Fig. \ref{lag_vs_wavelength} are total errors. Since the variability
at different radio bands is usually correlated and the observing times
were approximately the same in most cases, it follows that the errors
in our lag estimates for the different bands are also correlated.
Consequently, the lag uncertainty for one band relative to the other
bands is smaller than implied by the error bars. This holds for all
radio bands with the exception of the sub-mm APEX observations at
0.8\,mm that were performed not simultaneous to the cm/mm bands and
also include a slightly different sample and a smaller number of
sources. Although still consistent with the 2\,mm band lag given our
uncertainties, this likely also explains the slightly higher lag we
obtain at 0.8\,mm ($11\pm6$\,days) compared to 2\,mm ($7\pm9$\,days).

\section{Discussion}\label{discussion}

\subsection{Correlated radio/$\gamma$-ray variability and shocks}

Several detailed previous and ongoing studies provide support for
shocks \citep[e.g.][]{1985ApJ...298..114M,2011MmSAI..82..104T} as the
origin of the observed radio variability in cm/mm band blazar light
curves. Both detailed individual source and flare studies in the time
and/or spectral domain
\citep*[e.g.][]{1985ApJ...298..114M,2000A&A...361..850T,
  2011A&A...531A..95F,2013MNRAS.428.2418O,2013A&A...552A..11R} as well
as studies of larger samples and/or many individual flares
\citep[e.g.][]{1992A&A...254...71V,1994ApJ...437...91S,2008A&A...485...51H}
often show an overall good agreement of the observed flare signatures
with a shock-in-jet scenario. In particular, the multi-frequency radio
variability/flare amplitudes and time lags as well as the observed
spectral evolution seen in the F-GAMMA radio data sets (i.e. the data
also used in the present study) are in good agreement with the three
stages of shock evolution for most of the sources \citep[see][Fuhrmann
et al., Angelakis et al. in
prep.]{2011arXiv1111.6992A,fuhrmann2013,nestoras2013,2013MNRAS.428.2418O,2013A&A...552A..11R}.
Consequently, given the significant and strong radio/$\gamma$-ray
correlations presented in Sect. \ref{stacked_DCCF_results}, we
conclude that the bulk $\gamma$-ray emission/variability is likely
connected to the same shocked radio features. Those are first
appearing and evolving in the innermost, ultra-compact VLBI core
region and subsequently moving downstream the jet at pc scales with
apparent superluminal speeds as seen in VLBI images. This is in line
with similar conclusions of previous studies
\citep[e.g.][]{2003ApJ...590...95L,2011A&A...532A.146L}.

Assuming a leptonic emission scenario, the bulk $\gamma$-ray emission
would then be produced in the shocks by IC up-scattering of ambient
photons from the accretion disk, BLR, dusty torus and/or jet
(Synchrotron Self-Compton) depending on the shock location and
available ambient photon fields and their radiation energy densities
\citep[e.g.][]{2009MNRAS.397..985G}.

\subsection{Radio opacity and the location of the $\gamma$-ray emitting region}

Flare time delays in multi-frequency radio light curves are often
observed and commonly related to optical depth effects and travel-time
along the jet. The flare emission onsets and maxima appear typically
first at the highest radio frequencies, i.e. in the ``mm VLBI
core''. While subsequently moving downstream the jet and expanding
adiabatically, they become observable and peak at successively lower
frequencies where the optical depth of synchrotron self-absorption
(SSA) $\tau_{\mathrm{s}}$ (depending on the magnetic field and
electron energy density) decreases to about unity at the given radio
frequency.

Furthermore, radio opacity is also commonly observed as
frequency-dependent ``core shifts'' in multi-frequency VLBI images
\citep[e.g.][]{1998A&A...330...79L,2008A&A...483..759K,2011A&A...532A..38S}
with the core being identified as the most compact jet feature near
the apparent base of the jet, and the surface at which the optical
depth is $\approx$\,1 in a continuous jet flow
\citep[][]{1979ApJ...232...34B}. Assuming a conical jet geometry, the
VLBI core absolute positions shift according to
$r_{\mathrm{core}}\propto\nu^{-1/k_{\mathrm{r}}}$ where
$k_{\mathrm{r}}=[(3-2\alpha)m+2n-2]/(5-2\alpha)$.  Here $m$ and $n$
denote the power law exponents of the radial distance $r$-dependence
of the magnetic field $B(r)\propto r^{-m}$ and the electron density
$N(r)\propto r^{-n}$, and $\alpha$ denotes the optically thin spectral
index \citep[see][]{1998A&A...330...79L}, respectively.  For
SSA-dominated opacity and equipartition between jet particle and
magnetic-field energy density, $k_{\mathrm{r}}$ is 1 independent of
the spectral index for the choice $m=1$ and $n=2$
\citep[][]{1981ApJ...243..700K}.  In the presence of jet external
density and pressure gradients and/or foreground free-free absorption
(e.g.  due to BLR clouds), $k_{\mathrm{r}}$ becomes $>$\,1
\citep[e.g.][]{1998A&A...330...79L}.  However, if the observed radio
flare time lags are also due to opacity effects and the flare travels
at a constant speed, we expect $\tau\propto\nu^{-1/k_{\mathrm{r}}}$,
accordingly.  In this framework we examine the obtained
radio/$\gamma$-ray time lag frequency dependence shown in Fig.
\ref{lag_vs_wavelength}. The data are well described (reduced
$\chi^{2}<1$) by a power law function of the form
$\tau_{\mathrm{r,\gamma}}(\nu)=A+B\,\nu^{-1}$ (see least-square fits
in Fig.  \ref{lag_vs_wavelength}), i.e.  in good agreement with
SSA-dominated opacity and equipartition ($k_{\mathrm{r}}\simeq 1$).

Consequently, our results suggest a scenario where the $\gamma$ rays
escape instantaneously from the origin of the disturbance, whereas the
radio emission from the same region becomes optically thin at
progressively later times, i.e. flare maxima are observed successively
delayed w.r.t. the $\gamma$-ray emission.  A decreasing but still
positive lag towards mm wavelength with e.g.
$\tau_{\mathrm{3\,mm,\gamma}}=12\pm8$\,days then strongly suggests
that the bulk $\gamma$-ray production region is located inside and
even upstream of the 3\,mm core region. Towards higher frequencies the
situation is less clear given the small lags and large
uncertainties. At 2\,mm wavelengths, for instance, the estimated lag
is consistent with zero given our measurement uncertainties, possibly
even indicating co-spatial/contemporaneous emitting regions.

Using the average time lags we can estimate mean relative spatial
offsets $\Delta r_{\mathrm{r,\gamma}}$ between the region of
$\gamma$-ray peak emission and the ``$\tau=1$ photosphere'' of the
various radio bands, given by
\begin{equation}
\Delta r_{\mathrm{r,\gamma}}=\frac{\beta_{\mathrm{app}}c\,\tau^{\mathrm{source}}_{\mathrm{r,\gamma}}}{\sin\theta},
\end{equation} 
with $\theta$, $\beta_{\mathrm{app}}$ and
$\tau^{\mathrm{source}}_{\mathrm{r,\gamma}}$ the jet viewing angle,
apparent jet speed (assumed to be constant) and the source frame
radio/$\gamma$-ray time delay as obtained in
Sect. \ref{stacked_DCCF_results}, respectively.

Apparent jet speed measurements from the VLBI literature
\citep[e.g.][]{2009AJ....138.1874L} are available for 42 sources in
our sample. We note that the speed measurements used here are
collected from non-contemporaneous VLBI observations. Recent studies
show, however, that the dispersion of apparent speeds in a given jet
is moderate and jet speeds cluster around a characteristic value
\citep[][]{2013AJ....146..120L}. Viewing angles have been estimated
for these sources based on their $\beta_{\mathrm{app}}$ and
variability Doppler factors calculated from the F-GAMMA radio data at
20\,mm wavelength \citep[][Angelakis et al. in
prep.]{fuhrmann2013,nestoras2013}.  Using the mean values of $\theta$
(6.5$^{\circ}$) and $\beta_{\mathrm{app}}$ (17.5\,c) for the 42
sources we obtain mean, de-projected radio/$\gamma$-ray distances
decreasing from $9.8\pm3.0$\,pc at 110\,mm to $0.9\pm1.1$\,pc and
$1.4\pm0.8$\,pc at 2 and 0.8\,mm, respectively. The value of
$5.1\pm1.3$\,pc obtained at 20\,mm is furthermore in good agreement
with $\sim$\,7\,pc obtained by \citet{2010ApJ...722L...7P} based on a
comparable analysis using MOJAVE core flux densities at 20\,mm.  For
individual sources showing significant 3\,mm/$\gamma$-ray correlations
we provide estimates of $\Delta r_{\mathrm{3mm,\gamma}}$ in Table
\ref{single_source_results}. The obtained values range between 0.2 and
15.9\,pc.

\subsection{Application to 3C\,454.3}
The good agreement of our estimated time lags with SSA-dominated
opacity effects allows us to further constrain the location of the
$\gamma$-ray emitting region for individual sources in our sample. To
do so, we focus in the following on one particular source of Table
\ref{single_source_results}, quasar J2253+1608 (hereafter 3C\,454.3),
and combine our previous results ($\Delta r_{\mathrm{3mm,\gamma}}$)
with SSA arguments and a radio/radio time lag analysis to obtain a
lower limit for the distance of the $\gamma$-ray emitting region to
the SMBH in this source.

Using the same DCCF analysis as in Sect. \ref{DCCF_analysis} to also
obtain radio/radio time lags $\tau_{\mathrm{r,r}}$ for all radio
frequency combinations, we can estimate $k_{\mathrm{r}}$ and ``time
lag core shifts'' $\Delta r_{\mathrm{mas}}$ \citep[see
also][]{2006A&A...456..105B,2011MNRAS.415.1631K} for 3C\,454.3. The
knowledge of $k_{\mathrm r}$ (usually obtained from VLBI core shifts
at several frequencies) allows us to estimate the absolute distance of
the radio $\tau=1$ surface at a given frequency from the footpoint of
the jet (the ``jet-base'' or ``nozzle'') according to
$r_{\mathrm{base,\nu}}=\Omega_{\mathrm{r,\nu}}(\nu^{1/k_{\mathrm{r}}}\sin\theta)^{-1}$.
Here, $\Omega_{\mathrm{r,\nu}}\propto\Delta
r_{\mathrm{mas}}(\nu_{2}^{1/k_{\mathrm{r}}}-\nu_{1}^{1/k_{\mathrm{r}}})^{-1}$
is a measure of the position offset at $\nu_{1}$, $\nu_{2}$ with
$\nu_{2} > \nu_{1}$ \citep[see e.g.][]{1998A&A...330...79L}. $\Delta
r_{\mathrm{mas}}$ denotes the angular offset $\Delta
r_{\mathrm{mas}}=\mu\cdot\tau_{\mathrm{r,r}}$ with the VLBI jet proper
motion $\mu=0.3$\,mas/yr observed for 3C\,454.3
\citep[][]{2009AJ....138.1874L}. At 3\,mm wavelength we then obtain an
absolute distance of the ``$\tau=1$ surface'' from the jet-base
$r_{\mathrm{base,3mm}}$ of 1.8--2.6\,pc.

A distance estimate of the $\gamma$-ray emitting region from the
jet-base $r_{\mathrm{base,\gamma}}$ is then obtained using our
previously estimated $\Delta r_{\mathrm{r,\gamma}}$:
$r_{\mathrm{base,\gamma}}=r_{\mathrm{base,\nu}}-\Delta
r_{\mathrm{r,\gamma}}$, whereas the distance to the central SMBH is
given by
$r_{\mathrm{BH,\gamma}}=r_{\mathrm{BH,base}}+r_{\mathrm{base,\gamma}}$
with $r_{\mathrm{BH,base}}$ the distance between the BH and the
jet-base.  Using $\Delta r_{\mathrm{3mm,\gamma}}=1$\,pc obtained for
3C\,454.3 (Table \ref{single_source_results}) we estimate a distance
$r_{\mathrm{base,\gamma}}$ of 0.8--1.6\,pc.  Since the distance
$r_{\mathrm{BH,base}}$ is unknown (though likely small or negligible),
the obtained value provides a lower limit for the distance of the
$\gamma$-ray emitting region to the central SMBH in 3C\,454.3. Given
the typical BLR radii of $\lesssim$\,1\,pc observed in AGN and an
estimated value of $\sim$\,0.2\,pc for the bulk BLR material in
3C\,454.3 \citep[][]{2011MNRAS.410..368B}, our analysis of 3C\,454.3
reveals: (i) a 3\,mm $\tau=1$ surface (``core'') well outside the
canonical BLR, (ii) a location of the $\gamma$-ray emitting region
upstream of the 3\,mm core and at the outer BLR edge/extension or
beyond. We note that the BLR of 3C\,454.3 may be stratified and extend
well beyond the canonical radius of $\sim$\,0.2\,pc obtained from
scaling relations.  A more extended, pc-scale structure in e.g. an
``outflowing BLR'' scenario appears reasonable \citep[see
also][]{2011A&A...532A.146L}. It is interesting to link our results to
the recent findings of \citet{2013ApJ...763L..36L}. These authors find
BLR emission line variability in 3C\,454.3 being powered by the
non-thermal continuum emission of a new jet component traversing
through the mm-band radio core implying that the latter is surrounded
by BLR material. Given our findings of the ``mm-core'' being at
$\sim$\,2--3\,pc from the jet-base well outside the canonical BLR
radius thus supports the presence of BLR material even out to pc-scale
distances from the SMBH.

For a $\gamma$-ray emission location as obtained above, a leptonic
scenario suggests either the BLR and its extension, the pc-scale dusty
torus and/or the jet as main sources of seed photons for IC
up-scattering and bulk GeV photon production in 3C\,454.3.  A
$\gamma$-ray location with weak accretion disk radiation but still a
rich emission-line environment, i.e. the outer BLR, is in good
agreement with recent modeling of the observed broad-band SEDs and
$\gamma$-ray spectral breaks in this source using IC scattering of BLR
photons and energy densities near equipartition
\citep[][]{2013ApJ...771L...4C}.

\section{Summary and conclusions}\label{conclusions}
We have presented first results of a detailed cross-correlation
analysis between radio (cm, mm and sub-mm wavelengths of the F-GAMMA
program) and $\gamma$-ray variability in the $\sim$\,3.5 year light
curves of 54 {\it Fermi}-bright blazars. Our results for the studied
sample can be summarized as follows:

\begin{enumerate}
\renewcommand{\theenumi}{(\arabic{enumi})}
\item The 3.5 year light curves often display strong outbursts (time
  scales of months) and extended periods of activity (months to 1--2
  years) at both radio and $\gamma$ rays, whereas the $\gamma$-ray
  variability usually appears to be more rapid.

\item In order to increase the significance and sensitivity for
  correlations, a DCCF stacking analysis was performed using the whole
  sample and a new method to estimate correlation significances and
  chance correlations via a ``mixed source correlation'' method. For
  the latter analysis, we used a total of 131 $\gamma$-ray light
  curves including additional 77 reference blazars. This yields for
  the first time strong, statistically significant multi-band radio
  (11\,cm to 0.8\,mm) and $\gamma$-ray correlations.  The radio
  emission is typically lagging the $\gamma$ rays with sample average
  time lags ranging between 76\,$\pm$\,23 and 7\,$\pm$\,9\,days,
  systematically decreasing from the longer cm wavelengths to the
  mm/sub-mm bands.

\item The radio/$\gamma$-ray delay frequency dependence is well
  described by a power law
  $\tau_{\mathrm{r,\gamma}}(\nu)\propto\nu^{-1}$, as expected for
  synchrotron self-absorption (SSA) dominated opacity effects (with
  $\tau\propto\nu^{1/k_{{\mathrm r}}}$, $k_{{\mathrm r}}\simeq 1$).

\item Although the time lag rapidly decreases towards shorter
  wavelengths, a still positive delay at 3\,mm with
  $\tau_{\mathrm{3\,mm,\gamma}}=12\pm8$\,days suggests that the bulk
  $\gamma$-ray emission is coming from inside or even upstream of the
  (optically thick) 3\,mm-core region.

\item The mean spatial distances between the region of $\gamma$-ray
  peak emission and the radio ``$\tau=1$ photosphere'' are found to
  decrease from $9.8\pm3.0$\,pc at 110\,mm to $0.9\pm1.1$\,pc and
  $1.4\pm0.8$\,pc at 2 and 0.8\,mm wavelength, respectively.

\item Previous studies have shown that the multi-frequency radio
  variability observed in our sample is in overall good agreement with
  shocks and their three-stage evolution. Given the strong
  radio/$\gamma$-ray correlations presented here, we thus conclude
  that the enhanced, bulk $\gamma$-ray emission is likely also
  connected to these shocked jet structures.

\item We obtain 3\,mm/$\gamma$-ray correlations for 9 individual
  sources at a significance level where we expect one occurring by
  chance (chance probability: $4\times 10^{-6}$). These sources
  exhibit 3\,mm/$\gamma$-ray time lags $\tau_{{\mathrm 3mm,\gamma}}$
  in the source frame ranging between $-4\pm10$ and
  $93\pm16$\,days. No significant case of radio leading $\gamma$ rays
  is found.

\item The observed opacity/SSA effects allow us to further constrain the
  location of the $\gamma$-ray emitting region in individual sources
  using a new method which combines the radio/$\gamma$-ray as well as
  radio/radio time lags. Together with VLBI proper motion measurements
  and assuming a conical jet, ''time lag core shifts'' then reveal the
  absolute, de-projected distance of the bulk $\gamma$-ray emitting region
  from the jet-base. Applied to 3C\,454.3 we consequently obtain a
  lower limit for the $\gamma$-ray distance to the SMBH of 0.8--1.6\,pc.

\item For typical bulk BLR radii of $\lesssim$\,1\,pc observed in AGN
  and a value of $\sim$\,0.2\,pc obtained for 3C\,454.3, we place the
  $\gamma$-ray emitting region in this source at the outer edge or
  beyond the BLR. Our finding of the $\tau=1$ surface at 3\,mm being
  at $\sim$\,2--3\,pc from the jet-base (i.e. well outside the
  canonical BLR) together with recent findings of
  \citet{2013ApJ...763L..36L} suggests that BLR material in 3C\,454.3
  extends to several pc distances from the SMBH.

\end{enumerate}

Our overall findings suggest a scenario where the bulk light curve
flare emission is produced in shocks moving down the jet, whereas the
$\gamma$ rays are escaping instantaneously from the shocked jet region
and the optically thin radio emission from the same region reaches the
observer successively delayed due to opacity effects and travel-time
along the jet. The current low detection rate of significant
single-source correlations clearly demonstrates the need for longer
data trains and a correspondingly better ``event statistic'' to study
the correlation properties and $\gamma$-ray location for a larger
number of individual sources in detail. Our stacking analysis will
furthermore enable more detailed studies of the radio/$\gamma$-ray
correlation properties of the sample exploring possible differences of
the correlation behavior between different source classes (e.g. FSRQs
vs. BL Lacs) and testing dependencies on different physical parameters
such as black hole mass, BLR size, luminosity, jet opening angle and
Doppler factor (Fuhrmann et al., Larsson et al.  in prep.).

We stress that the overall situation is complex. The strong and
significant multi-band correlations presented here are statistical in
nature and often no simple, detailed one-to-one correlation of single
radio and $\gamma$-ray flares is observed. In addition, our
correlation method and data sets are mostly sensitive to the maxima
and minima of the most prominent, long-term variability/flares in the
studied sample. The radio/$\gamma$-ray correlation properties of the
more rapid $\gamma$-ray flares often observed in these sources on time
scales of $\lesssim$\,hours or days to a few weeks (and not resolved
with our data sets and analysis) may be different.  These events may
be produced also at different locations. However, the very smooth and
continuous behavior of the observed time lag in agreement with SSA all
the way to the mm and sub-mm bands may provide some evidence against
the ``43\,GHz standing shock and turbulent extreme multi-zone
scenario'' \citep[e.g.][]{2010arXiv1005.5551M,2013arXiv1304.2064M}.
Future, more detailed studies of single sources will shed further
light on this topic.

Neither our stacking method nor single-source results provide strong
evidence for cases of radio leading $\gamma$ rays. This demonstrates
the limited predictive power of radio flares to reliably trigger {\it
  Fermi} observations of flaring $\gamma$-ray sources (whereas the
detection of a $\gamma$-ray flare by {\it Fermi} likely signals an
impending high state at radio bands).  This statement holds unless
radio (mm/sub-mm) flare onsets occur simultaneously with or even
before $\gamma$-ray flare onsets, which is unclear at the moment. This
needs to be addressed in future studies.

\section*{Acknowledgments}
This research is based on observations with the 100-m telescope of the
MPIfR (Max-Planck-Institut f\"ur Radioastronomie) at Effelsberg. This
work has made use of observations with the IRAM 30-m telescope. IRAM
is supported by INSU/CNRS (France), MPG (Germany) and IGN (Spain).
Part of this work was supported by the COST Action MP0905 "Black Holes
in a Violent Universe". Stefan Larsson also acknowledge support by a
grant from the Royal Swedish Academy Crafoord Foundation.  The
\textit{Fermi} LAT Collaboration acknowledges generous ongoing support
from a number of agencies and institutes that have supported both the
development and the operation of the LAT as well as scientific data
analysis.  These include the National Aeronautics and Space
Administration and the Department of Energy in the United States, the
Commissariat \`a l'Energie Atomique and the Centre National de la
Recherche Scientifique / Institut National de Physique Nucl\'eaire et
de Physique des Particules in France, the Agenzia Spaziale Italiana
and the Istituto Nazionale di Fisica Nucleare in Italy, the Ministry
of Education, Culture, Sports, Science and Technology (MEXT), High
Energy Accelerator Research Organization (KEK) and Japan Aerospace
Exploration Agency (JAXA) in Japan, and the K.~A.~Wallenberg
Foundation, the Swedish Research Council and the Swedish National
Space Board in Sweden.  Additional support for science analysis during
the operations phase is gratefully acknowledged from the Istituto
Nazionale di Astrofisica in Italy and the Centre National d'\'Etudes
Spatiales in France.  I. Nestoras is supported for this research
through a stipend from the International Max Planck Research School
(IMPRS) for Astronomy and Astrophysics at the Universities of Bonn and
Cologne.

We thank Charles Dermer, Andrei Lobanov and the referee for valuable
comments and suggestions.

\bibliographystyle{mn2e}
\bibliography{references}

\label{lastpage}

\end{document}